\documentclass[%
superscriptaddress,
nofootinbib,
nobibnotes,
amsmath,amssymb,
aps,
floatfix,
]{revtex4-2}

\usepackage{placeins}
\usepackage{booktabs}
\usepackage{dcolumn}
\usepackage{array}
\usepackage{longtable}
\usepackage{float}
\usepackage{hyphenat}
\usepackage{natbib}
\usepackage{hyperref}

\usepackage{amsmath,amsfonts,amsthm,amssymb}
\usepackage{bm,graphicx,graphics,color}
\usepackage{epsf,epsfig}
\usepackage[all]{xy}
\newcommand*{\B}[1]{\ifmmode\bm{#1}\else\textbf{#1}\fi}

\def\bx{\mathbf{x}}

\def\bv{\mathbf{v}}
\def\bV{\mathbf{V}}

\def\bp{\mathbf{p}}

\def\no{\nonumber}
\def\lb{\label}
\def\be{\begin{equation}}
\def\ee#1{\label{#1}\end{equation}}
\newcommand{\ben}{\begin{eqnarray}}
\newcommand{\een}{\end{eqnarray}}

\usepackage[english]{babel}

\usepackage[letterpaper,top=2cm,bottom=2cm,left=3cm,right=3cm,marginparwidth=1.75cm]{geometry}

\usepackage{amsmath}
\usepackage{graphicx}
\begin{document}

\title{Post-Newtonian non-equilibrium kinetic theory }%Non-equilibrium post-Newtonian hydrodynamics}

\author{Gilberto M. Kremer}
\email{kremer@fisica.ufpr.br}
\affiliation{Departamento de F\'{i}sica, Universidade Federal do Paran\'{a}, Curitiba 81531-980, Brazil}

\begin{abstract}
The post-Newtonian hydrodynamic equations for a non-perfect fluid are developed within the framework of a post-Newtonian Boltzmann equation.  The post-Newtonian components of the energy-momentum tensor  are determined by considering the relativistic Eckart decomposition for a viscous  and heat conducting fluid. From the relativistic  Grad distribution function its post-Newtonian expression is derived. The hydrodynamic equations for the mass density, mass-energy density and momentum density are determined from a post-Newtonian transfer equation and Grad's distribution function. In the non-relativistic limit the Newtonian hydrodynamic equations for mass, momentum and energy densities are recovered.
\end{abstract}


\maketitle
\section{Introduction}\lb{sec1}

  In 1938 Einstein, Infeld and Hoffmann \cite{Eins} proposed a method of successive approximations in powers of $1/c^2$ for the solution of Einstein's field equations which goes beyond the Newtonian gravitational theory. In the post-Newtonian theory  the components of the metric tensor in the order $1/c^n$ are determined from Einstein's field equations  once the  energy-momentum tensor in the order $1/c^{n-2}$ is known.  For a energy-momentum tensor described by a constitutive equation of a perfect fluid the hydrodynamic equations in the first post-Newtonian  approximation were determined by Chandrasekhar \cite{Ch1} and Weinberg \cite{Wein} while the corresponding ones in the second post-Newtonian approximation by Chandrasekhar and Nutku \cite{ChNu} (see also \cite{PGMK,GGKK}).
  
  A subject which is interesting to investigate is the inclusion of non-equilibrium quantities related with the viscous stress and the heat flux in the post-Newtonian  energy-momentum tensor of a non-perfect fluid. The role of the viscous stress in the post-Newtonian theory was first analysed in  \cite{JH}. 
  
  The aim of this work is to derive within the framework of  kinetic theory of gases the post-Newtonian hydrodynamic equations for a viscous and heat conducting fluid. We  call attention to the fact that the introduction of non-equilibrium quantities related with the viscous stress and heat flux does not change the determination of the components of the metric tensor from Einstein's field equations in the first post-Newtonian approximation.
  
  We start by determining the post-Newtonian expressions for the components of the energy-momentum tensor which for a non-perfect fluid is described by the Eckart decomposition \cite{Eck}.
  
  At equilibrium the one-particle distribution function for a relativistic gas is characterized by the Maxwell-J\"uttner distribution function (see e.g. \cite{CK}) and its post-Newtonian version was determined in \cite{KRW}.
  For non-equilibrium processes the one-particle distribution function may be characterized by Grad's distribution function \cite{Grad}. In the relativistic case  Grad's distribution function is a function of the fourteen fields of particle number density, four-velocity, absolute temperature, dynamic pressure, pressure deviator and heat flux (see e.g. \cite{CK}). In this work the post-Newtonian approximation of the relativistic Grad distribution function is derived which is used to determine the contributions of the viscous stress and heat flux to the components of the energy-momentum tensor.
  
  The Boltzmann equation in the first post-Newtonian approximation was derived in \cite{Ped,Rez} and its version in the second post-Newtonian approximation in \cite{PGMK,GGKK}. 
  Here from the first post-Newtonian Boltzmann equation the so-called Maxwell-Enskog transfer equation \cite{Max,Ens}  is derived for arbitrary macroscopic quantities which are associated with mean values of microscopic quantities.

 The hydrodynamic equations for the mass density, mass-energy density and momentum density are obtained from the transfer equation  by considering the rest mass and the post-Newtonian expressions for the components of the particle momentum four-vector  together with the post-Newtonian Grad distribution function. These hydrodynamic equations correspond to the ones that follow  from the conservation equations  of the particle four-flow and energy-momentum tensor in the post-Newtonian approximation. Without the relativistic corrections  the hydrodynamic equations recover the Newtonian hydrodynamic equations for the mass, momentum and energy densities of a non-perfect fluid.

The paper  is outlined as follows: in Section \ref{sec2} we determine the components of the post-Newtonian non-equilibrium energy momentum tensor in the Eckart decomposition. The  post-Newtonian Boltzmann equation, the transfer equation and Grad's distribution function are determined Sections \ref{sec3}. In Section \ref{sec4} the post-Newtonian hydrodynamic equations are derived and the conclusions of the work are stated in Section \ref{sec5}.  Here Latin indices take the values 1,2,3 and the Greek indices take the values 0,1,2,3. Furthermore, the indices of Cartesian tensors will be written as subscripts, the summation convention over repeated  indices will be assumed, the partial differentiation will be denoted by $\partial/\partial x^i$ and the covariant differentiation is denoted by a semicolon.

\section{Energy-momentum tensor}\lb{sec2}

The hydrodynamic equations of a relativistic fluid are determined by the conservation laws of the particle four-flow $N^\mu$ and energy-momentum tensor $T^{\mu\nu}$, namely
\ben\lb{ne0}
{N^{\mu}}_{;\mu}=0,\qquad {T^{\mu\nu}}_{;\nu}=0.
\een

The  identification of  the relativistic non-equilibrium quantities with the non-relativistic ones  is attained by  introducing the decompositions of the particle four-flow $N^\mu$ and energy-momentum tensor $T^{\mu\nu}$  with respect to the four-velocity $U^{\mu}$ (such that $U^\mu U_\mu=c^2$). In the literature there exist two representations for the non-equilibrium particle four-flow and energy-momentum tensor known as  the  Eckart \cite{Eck} and the Landau and Lifshitz \cite{LL1} decompositions. Both representations make use of the projector
\ben\lb{ne1}
\Delta^{\mu\nu}=g^{\mu\nu}-
\frac1{c^2} U^{\mu}U^{\nu},
\een
 which projects an arbitrary four-vector into another four-vector perpendicular to the four-velocity. The projector has the following properties
\ben\lb{ne2}
\Delta^{\mu\nu}U_\nu=0,\qquad \Delta^{\mu\nu}\Delta_{\nu\sigma}=
{\Delta^{\mu}}_{\sigma},
\quad
{\Delta^{\mu}}_{\nu}
\Delta^{\nu\sigma}=
\Delta^{\mu\sigma},
\quad
{\Delta^{\mu}}_{\mu}=3,
\een
and in a local Minkowski  rest frame  where $U^\mu=(c,\bf0)$  it reads $\Delta^{\mu\nu}={\rm diag}(0,-1,-1,-1)$. 

By using the projector one can introduce for an arbitrary four vector $A^\mu$  and tensor $A^{\mu\nu}$ the following representations
\ben\lb{ne3a}
A^{(\mu)}={\Delta^{\mu}}_{\nu} A^{\nu},
\qquad
A^{ (\mu\nu)}=\frac12
\left(
{\Delta^{\mu}}_{\sigma}
{\Delta^{\nu}}_{\tau}
+{\Delta^{\nu}}_{\sigma}
{\Delta^{\mu}}_{\tau}
\right)A^{\sigma\tau},
\\\lb{ne3c}
A^{ [\mu\nu]}=\frac12
\left(
{\Delta^{\mu}}_{\sigma}
{\Delta^{\nu}}_{\tau}
-{\Delta^{\nu}}_{\sigma}
{\Delta^{\mu}}_{\tau}
\right)A^{\sigma\tau},
\qquad
A^{\langle\mu\nu\rangle}
=A^{ (\mu\nu)}
-\frac13\Delta^{\mu\nu}
\Delta_{\sigma\tau}A^{(\sigma\tau)},
\een
which are associated with  a four-vector, a symmetric tensor,
an antisymmetric tensor and  a symmetric traceless tensor that have only the spatial components in a local Minkowski rest frame, respectively. Note that  $\Delta_{ \mu\nu}
A^{\langle\mu\nu\rangle}=0$ hold.

Here we shall use the Eckart decomposition
where the particle four-flow $N^{\mu}$ and
the energy-momentum tensor
$T^{\mu\nu}$ for a viscous heat conducting fluid are
represented  as:
\ben\lb{ne4a}
N^{\mu}=nU^{\mu},
\qquad
T^{\mu\nu}=p^{\langle \mu\nu\rangle}-\left(p+\varpi\right)
\Delta^{\mu\nu}
+\frac\epsilon{c^2}
U^{\mu}U^{\nu}
+\frac1{c^2}
\bigg(U^{\mu}q^{(\nu)}
+U^{\nu}q^{(\mu)}\bigg).
\een
 The quantities $n$,
$p^{\langle \mu\nu\rangle}$,
$p$, $\varpi$, $q^{(\mu)}$ and $\epsilon$ introduced by the above decompositions are identified as follows:
\ben\lb{ne5a}
n=\frac1{c^2} N^{\mu} U_{\mu}
\hbox{--  particle number density,}
\\\lb{ne5b}
p^{\langle \mu\nu\rangle}
=\left(
{\Delta^{\mu}}_{\sigma}
{\Delta^{\nu}}_{\tau}
-\frac13
\Delta^{\mu\nu}\Delta_{\sigma\tau}\right)
T^{\sigma\tau} \hbox{-- pressure deviator,}
\\\lb{ne5c}
\left (p+\varpi\right)
=-\frac13 \Delta_{\mu\nu}T^{\mu\nu}
\hbox{-- hydrostatic  +
dynamic pressures,}
\\\lb{ne5d}
q^{(\mu)}={\Delta^{\mu}}_{\nu}U_{\sigma}
T^{\nu\sigma}
\hbox{ -- heat flux,}
\\\lb{ne5e}
\epsilon=\frac1{c^2} U_{\mu}
T^{\mu\nu}U_{\nu}
\hbox{ -- energy density.}
\een
 The
hydrostatic pressure $p$ and
the energy density $\epsilon$ refer to equilibrium quantities of the energy-momentum tensor and the dynamic pressure $\varpi $ is the non-equilibrium part of its trace. Furthermore, the energy density $\epsilon=\rho c^2(1+\varepsilon/c^2)$ is a sum of two terms one related with the mass density $\rho=mn$ -- where $m$ is the particle rest mass -- and the other with the internal energy density $\rho\varepsilon$.

%\section{Post-Newtonian approximation}

The post-Newtonian approximations  refer to the solutions of Einstein's field equations  in successive powers of $1/c^2$  \cite{Eins,Ch1,Wein}.  Here we shall follow the work of Chandrasekhar \cite{Ch1} and write  the first post-Newtonian approximation for the components of the metric tensor  as
\ben\lb{ne6a}
g_{00}=1-\frac{2U}{c^2}+\frac2{c^4}\left(U^2-2\Phi\right),\qquad
g_{0i}=\frac{\Pi_i}{c^3},\qquad g_{ij}=-\left(1+\frac{2U}{c^2}\right)\delta_{ij},
\een
where the scalar gravitational potentials  $U$, $\Phi$ and the vector gravitational potential  $\Pi_i$  in the above equations   satisfy the Poisson equations  
\ben\lb{ne6b}
\nabla^2U=-4\pi G\rho,\qquad\nabla^2\Phi=-4\pi G\rho\left(V^2+U+\frac\varepsilon2+\frac{3p}{2\rho}\right),
\qquad
\nabla^2\Pi_i=-16\pi G\rho V_i+\frac{\partial^2U}{\partial t\partial x^i}.
\een
Here $G$ is the universal gravitational constant and $\bV$ the three hydrodynamic velocity.

The components of the four-velocity up to the order $1/c^4$ are
\ben\lb{ne6c}
U^0=c\left[1+\frac1{c^2}\left(\frac{V^2}2+U\right)
+\frac1{c^4}\left(\frac{3V^4}8+\frac{5U V^2}2+\frac{U^2}2+2\Phi-\Pi_iV_i\right)\right],
\qquad
U^i=\frac{V_iU^0}c.
\een

The components of the projector in the first post-Newtonian approximation follows from (\ref{ne6a}) and (\ref{ne6c}) and read
\ben\lb{ne7a}
\Delta^{00}=-\frac{V^2}{c^2}+\mathcal{O}(c^{-4}),
\quad
\Delta^{0i}=-\frac{V_i}{c}+\mathcal{O}(c^{-3}),\qquad
%\\\lb{ne7b}
\Delta^{ij}=-\left(1-\frac{2U}{c^2}\right)\delta_{ij}-\frac{V_iV_j}{c^2}+\mathcal{O}(c^{-4}).
\een
Here $\mathcal{O}(c^{-n})$ denotes the  order of the $n$th inverse power of the light speed.

Let us first analyse the components of the pressure deviator $p^{\langle \mu\nu\rangle}$.  From the relationship $U_\mu   p^{\langle \mu\nu\rangle}=g_{\nu\sigma}U^\sigma p^{\langle \mu\nu\rangle}=0$ we have that
\ben\lb{ne8a}
\left(g_{00}U^0+g_{0j}U^j\right) p^{\langle0i\rangle}+\left(g_{0j}U^0+g_{jk}U^k\right)p^{\langle ij\rangle}=0,\\\lb{ne8}
\left(g_{00}U^0+g_{0j}U^j\right) p^{\langle00\rangle}+\left(g_{0j}U^0+g_{jk}U^k\right)p^{\langle 0j\rangle}=0,\een
which by considering (\ref{ne6a}) and (\ref{ne6c}) imply the following relations for the time and space-time components of the pressure deviator
\ben\lb{ne8b}
p^{\langle00\rangle}=p^{\langle ij\rangle}\frac{V_iV_j}{c^2}+\mathcal{O}(c^{-4}),\qquad p^{\langle0i\rangle}=p^{\langle ij\rangle}\frac{V_j}c+\mathcal{O}(c^{-3}).
\een

In order to fulfill the  traceless condition of the pressure deviator up to the order of the first post-Newtonian approximation we represent the spatial  components  of the pressure  deviator as
\ben\lb{ne8c}
p^{\langle ij\rangle}=\mathfrak{p}_{ij}+\frac1{2c^2}\left(\mathfrak{p}_{ik}V_kV_j+\mathfrak{p}_{jk}V_kV_i\right)+\mathcal{O}(c^{-4}).
\een
It is easy to verify that the above  representation fulfills the traceless condition $\Delta_{\mu\nu}p^{\langle \mu\nu\rangle}=g_{\mu\nu}p^{\langle \mu\nu\rangle}=0$ up to the first post-Newtonian approximation. Above we have introduced the   non-relativistic  pressure deviator
\ben\lb{ne8d}
\mathfrak{p}_{ij}=p_{ij}-\frac{p_{rr}}3\delta_{ij},\qquad\hbox{such that}\qquad \delta_{ij}\mathfrak{p}_{ij}=0.
\een

%The pressure deviator components (\ref{ne8b}) can be rewritten thanks to (\ref{ne8c}) as
%\ben\lb{ne8e}
%p^{\langle00\rangle}=\mathfrak{p}_{ij}\frac{V_iV_j}{c^2}+\mathcal{O}(c^{-4}),\qquad p^{\langle0i\rangle}=\mathfrak{p}_{ij}\frac{V_j}c+\mathcal{O}(c^{-3}).
%\een

For the components of the heat flux we make use of  the relationship $U_\mu q^{(\mu)}=g_{\mu\nu}U^\nu q^{(\mu)}=0$ and get  that its time component  becomes
\ben\lb{ne8f}
 q^{(0)}=\mathfrak{q}_i\frac{V_i}c+\mathcal{O}(c^{-3}),
\een
where $q^{(i)}=\mathfrak{q}_i$ is the non-relativistic heat flux vector.

The non-relativistic pressure deviator $\mathfrak{p}_{ij}$ and the heat flux vector $\mathfrak{q}_i$ vanish at equilibrium and in the non-relativistic limiting case  we have 
$p^{\langle ij\rangle}=\mathfrak{p}_{ij},$ $p^{\langle00\rangle}=0,$
$p^{\langle0i\rangle}=0,$  $q^{(i)}=\mathfrak{q}_i$ and $q^{(0)}=0$.

 The energy-momentum tensor components can be obtained now  from (\ref{ne4a}) together with  (\ref{ne7a}),   (\ref{ne8b}) -- (\ref{ne8f}), yielding
\ben\lb{ne9a}
&&T^{00}= \rho c^2\left[1+\left(V^2+\varepsilon+2U\right)\right],
\\\lb{ne9b}
&&T^{i0}=\rho cV_i\left[1+\frac1{c^2}\left(V^2+2U+\varepsilon+\frac{p}\rho\right)\right]
+\frac{\mathfrak{p}_{ij}V_j}{ c}+\frac{\mathfrak{q}_i}{c},
\\\no
&&T^{ij}=\rho V_iV_j+p\delta_{ij}+\mathfrak{p}_{ij}+\frac\rho{c^2}\left(V^2+2U+\varepsilon+\frac{p}\rho\right)V_iV_j -2\frac{p}{c^2} U\delta_{ij}+\frac1{c^2}\left(\mathfrak{q}_iV_j+\mathfrak{q}_jV_i\right)
\\\lb{ne9c}
&&\qquad+\frac1{2c^2}\left(\mathfrak{p}_{ik}V_kV_j+\mathfrak{p}_{jk}V_kV_i\right).
\een
In the above equations the dynamic pressure $\varpi$ was not taken into account, since for rarefied monatomic gases its constitutive equation is proportional to the velocity divergent and the coefficient of proportionality -- the bulk or volume viscosity -- is of order of $1/c^4$ (see \cite{CK}).

Without the non-equilibrium pressure deviator  $\mathfrak{p}_{ij}$ and heat flux vector $\mathfrak{q}_{i}$   the components of the energy-momentum tensor (\ref{ne9a}) -- (\ref{ne9c}) reduce to the well known expressions in the literature \cite{Ch1,Wein} for perfect fluids.

\section{Post-Newtonian kinetic theory}\lb{sec3}

In  kinetic theory of gases the space-time evolution of the one-particle distribution function $f(\bx,\bp,t)$ in the phase space spanned by the  spatial coordinates $\bx$ and momentum  $\bp$ of the particles  is governed by the Boltzmann equation. The first post-Newtonian approximation of the Boltzmann equation  read \cite{Ped,PGMK,GGKK}
\ben\no
\bigg[\frac{\partial f}{\partial t}+v_i\frac{\partial f}{\partial x^i}\bigg]\bigg[1+\frac1{c^2}\bigg(\frac{v^2}{2}+U\bigg)\bigg]+\frac{\partial U}{\partial x^i}\frac{\partial f}{\partial v_i}-\frac1{c^2}\bigg[4v_iv_j\frac{\partial U}{\partial x^j}+3v_i\frac{\partial U}{\partial t}
\\\lb{ne10a}
-\left(\frac{3v^2}2-3U\right)\frac{\partial U}{\partial x^i}
-2\frac{\partial \Phi}{\partial x^i}-\frac{\partial \Pi_i}{\partial t}-v_j\bigg(\frac{\partial \Pi_i}{\partial x^j}-\frac{\partial \Pi_j}{\partial x^i}\bigg)\bigg]\frac{\partial f}{\partial v_i}=\mathcal{Q}(f,f).
\een
Above $\mathcal{Q}(f,f)$ is the collision operator of the Boltzmann equation, which takes into account the binary collisions of the particles and is represented by an integral of the product of two particle distribution functions at collision.

 Furthermore, the  energy-momentum tensor in  kinetic theory of gases is defined in terms of the one-particle distribution function by \cite{CK}
\ben\lb{ne10b}
T^{\mu\nu}=m^4c\int u^\mu u^\nu f\frac{\sqrt{-g}\,d^3 u}{u_0}.
\een
Here  $m$ is the particle rest mass, $u^\mu=p^\mu/m$  the particle four-velocity and ${\sqrt{-g}\,d^3 u}/{u_0}$ the invariant integration element. The components of the particle four-velocity in the post-Newtonian approximation reads
\ben\lb{ne10c}
u^0=c\bigg[1+\frac1{c^2}\bigg(\frac{v^2}2+U\bigg)+\frac1{c^4}\bigg(\frac{3v^4}8+\frac{5U v^2}2
+\frac{U^2}2+2\Phi-\Pi_iv_i\bigg)\bigg],\qquad u^i=v_i \frac{u^0}c,
\een
which has the same structure as (\ref{ne6c}), the difference being that the hydrodynamic $\bf V$ is substituted by the particle velocity $\bf v$.

The invariant integration element of the energy-momentum tensor (\ref{ne10b}) in the  first post-Newtonian approximation  was determined in \cite{KRW} and is given by
\ben\lb{ne10d}
\frac{\sqrt{-g}\, d^3 u}{u_0}=
\left\{1+\frac1{c^2}\left[2v^2+6U\right]\right\}\frac{d^3 v}c.
\een

At  equilibrium the collision operator of the Boltzmann equation vanishes and the one-particle distribution function becomes the Maxwell-J\"uttner distribution function (see e.g \cite{CK})
\ben\lb{ne11a}
f^{(0)}=\frac{n}{4\pi m^2ckTK_2(\zeta)}\exp
\left(-\frac{p^\mu U_\mu}{kT}\right).
\een
Above $T$ is the absolute temperature, $k$  the Boltzmann constant and    $\zeta=mc^2/kT$  a relativistic parameter which is given by  the ratio of the rest energy of the gas particles $mc^2$ and the thermal energy of the gas $kT$. In the ultra-relativistic limiting case $\zeta\ll1$  and in the non-relativistic limiting case $\zeta\gg1$. Furthermore,  $K_2(\zeta)$ denotes the modified Bessel function of the second kind.

The first post-Newtonian approximation of the Maxwell-J\"uttner distribution function (\ref{ne11a}) was determined in \cite{KRW} and its expression is
\ben\lb{ne11b}
f^{(0)}=\frac{ne^{-\frac{m\mathcal{V}^2}{2kT}}}{(2\pi mkT)^{\frac32}}
\Bigg\{ 1-\frac{1}{c^2}\bigg[\frac{15 kT}{8 m}+\frac{m (V_i\mathcal{V}_i)^2}{2 kT}+\frac{2 m U \mathcal{V}^2}{kT}+\frac{3 m \mathcal{V}^4}{8 kT}+\frac{m V^2 \mathcal{V}^2}{2 kT}+\frac{m (V_i\mathcal{V}_i) \mathcal{V}^2}{kT}\bigg]\Bigg\}.\;
\een
Here $\mathcal{V}_i=v_i-V_i$ is the so-called peculiar velocity which is the particle velocity  in the gas frame i.e., the difference of the particle velocity $v_i$ and the gas velocity $V_i$. 

In the non-relativistic kinetic theory of gases Grad's distribution function  plays an important role to describe the non-equilibrium behavior of gases not too far from a local equilibrium. Grad's moment method  \cite{Grad} was proposed in 1949 for a non-relativistic gas and takes into account the thirteen  moments of the one-particle distribution function: particle number density, hydrodynamic velocity, pressure tensor and heat flux vector.
For the case of a relativistic gas one has to include the field of the
dynamic pressure which is the trace of the pressure tensor in non-equilibrium and in this case  Grad's distribution function is described by fourteen basic fields. The relativistic
Grad's  distribution function in terms of the fourteen
fields of particle number density $n$, four-velocity $U^\mu$, absolute temperature $T$, dynamic pressure $\varpi$, pressure deviator $p^{\langle\mu\nu\rangle}$ and heat flux
$q^{(\mu)}$ is ( see e.g.\cite{CK})
\ben\no
f=f^{(0)}\Bigg\{1+\frac{q^{(\mu)}}{p}\frac{\zeta}{
\zeta+5\mathcal{G}-\mathcal{G}^2\zeta}\left[\frac{\mathcal{G} p_\mu}{mc^2} 
-\frac{U_{\nu}p^{\nu}p_{\mu}}{m^2c^4}\right]+\frac{p^{\langle\mu\nu\rangle}}p
\frac{\zeta p_{\mu}p_{\nu}}{2\mathcal{G} m^2c^2}
\\\no
+\frac{\varpi}{p}\frac{
1-5\mathcal{G}\zeta -\zeta^2+\mathcal{G}^2\zeta^2}{
20\mathcal{G}+3\zeta-13\mathcal{G}^2\zeta-2\mathcal{G}\zeta^2+2\mathcal{G}^3\zeta^2}
\Biggl[\frac{U_{\mu}p^{\mu}}{mc^2}
\frac{3\zeta(6\mathcal{G}+\zeta-\mathcal{G}^2\zeta)}{
1-5\mathcal{G}\zeta -\zeta^2+\mathcal{G}^2\zeta^2}
\\\lb{ne11c}
+\frac{\zeta U_{\mu}U_{\nu}p^{\mu}p^{\nu}}{m^2c^4}
+\frac{15\mathcal{G}+2\zeta-6\mathcal{G}^2\zeta+5\mathcal{G}\zeta^2+\zeta^3-\mathcal{G}^2\zeta^3}{
1-5\mathcal{G}\zeta -\zeta^2+\mathcal{G}^2\zeta^2}\Biggr]
\Bigg\},
\een
where $f^{(0)}$ is the Maxwell-J\"uttner distribution function (\ref{ne11a}) and $\mathcal{G}$ denotes the ratio of the Bessel functions $\mathcal{G}=K_3(\zeta)/K_2(\zeta)$.

The first post-Newtonian approximation of Grad's distribution function (\ref{ne11c}) is obtained by taking into account the expressions for the hydrodynamic four-velocity (\ref{ne6c}), particle four-velocity (\ref{ne10c}),  the asymptotic expression for the modified Bessel functions of second kind $K_n(\zeta)$ for large values of $\zeta\gg 1$ \cite{AbSt}
\ben
K_n(\zeta)=
\sqrt{\frac\pi{2\zeta}}\frac1{e^\zeta}
\left[ 1+\frac{4n^2-1}{8\zeta}
+\frac{(4n^2-1)(4n^2-9)}{2!(8\zeta)^2}
+\dots\right],
\een
and the relations for the components of the pressure deviator and heat flux, namely
\ben\lb{ne12a}
p^{\langle00\rangle}=p^{\langle ij\rangle}\frac{V_iV_j}{c^2}+\frac2{c^4}p^{\langle ij\rangle}V_j\left(4U V_i-\Pi_i\right)+\mathcal{O}(c^{-6}),\\ p^{\langle0i\rangle}=p^{\langle ij\rangle}\frac{V_j}c+\frac{p^{\langle ij\rangle}}{c^3}\left(4U V_j-\Pi_j\right)+\mathcal{O}(c^{-5}).
\\\lb{ne12b}
 q^{(0)}=\mathfrak{q}_i\frac{V_i}c+\frac{\mathfrak{q}_i}{c^3}\left(4U V_i-\Pi_i\right)+\mathcal{O}(c^{-5}).
\een

After a long calculation one can obtain the first post-Newtonian approximation of Grad's distribution function
\ben\no
&&f=f^{(0)}\Bigg\{1+
\left(\frac{m}{kT}\right)^2\frac{\mathcal{V}_i\mathcal{V}_j}{2\rho}\left\{\mathfrak{p}_{ij}\left[1+\frac1{c^2}\left(\frac{\mathcal{V}^2+V^2+2\mathcal{V}_iV_i}{2}+6U-\frac{5kT}{2m}\right)\right]+\frac{\mathfrak{p}_{ik}V_kV_j+\mathfrak{p}_{jk}V_kV_i}{2 c^2}\right\}
\\\no
&&\qquad\underline{+\frac{\varpi}\rho\left(\frac{mc}{kT}\right)^2\left(\frac{m\mathcal{V}^2}{kT}-\frac32-\frac{m^2\mathcal{V}^4}{10k^2T^2}\right)}-\frac{\mathcal{V}_i\mathfrak{q}_{i}}{\rho}\left(\frac{m}{kT}\right)^2\bigg\{\left(1-\frac{m\mathcal{V}^2}{5kT}\right)+\frac{1}{c^2}\bigg[\frac{4\mathcal{V}^2}{5}+\frac{{V}^2}{2}+3U
\\\lb{ne13a}
&&\qquad+\mathcal{V}_jV_j-\frac{3kT}{4m}-\frac{m}{20kT}\left(5\mathcal{V}^4+4(V_j\mathcal{V}_j)^2+6V^2\mathcal{V}^2+28\mathcal{V}^2U+12(V_j\mathcal{V}_j)\mathcal{V}^2\right)\bigg]\bigg\}
\Bigg\}.
\een
In the above equations $f^{(0)}$ is the first post-Newtonian approximation for the Maxwell-J\"uttner distribution function (\ref{ne11b}).

By neglecting the $1/c^2$ terms (\ref{ne13a}) reduces to well-known non-relativistic Grad's distribution function
\ben\lb{ne13b}
f=\frac{ne^{-\frac{m\mathcal{V}^2}{2kT}}}{(2\pi mkT)^{\frac32}}\Bigg\{1+
\left(\frac{m}{kT}\right)^2\frac{\mathcal{V}_i\mathcal{V}_j}{2\rho}\mathfrak{p}_{ij}
-\frac{\mathcal{V}_i\mathfrak{q}_{i}}{\rho}\left(\frac{m}{kT}\right)^2\left(1-\frac{m\mathcal{V}^2}{5kT}\right)
\Bigg\}.
\een

If we insert the first post-Newtonian approximation of Grad's distribution function (\ref{ne13a}) and the invariant element of integration (\ref{ne10d}) into the definition of the energy-momentum tensor (\ref{ne10b}) and integrate the resulting equation by making use of the integrals of the appendix, we get 
the components of the energy-momentum tensor (\ref{ne9a}) -- (\ref{ne9c}). For a complete identification one has to consider the equation of state $p=\rho kT/m$  and the equation for specific internal energy $\varepsilon=3kT/2m$ of a monatomic gas.

Note that we have taken into account the underlined term in (\ref{ne13a}) which refers to the dynamic pressure, but the contribution of this term to the energy-momentum tensor vanishes.

\section{Hydrodynamic equations}\lb{sec4}
The determination of the hydrodynamic equations  from a transfer equation derived from the Boltzmann equation is an old task in the literature of the kinetic theory of gases which goes back to the works of Maxwell \cite{Max} and Enskog \cite{Ens}. Here we shall determine the so-called Maxwell-Enskog transfer equation in the first post-Newtonian approximation. 

To this end we multiply the Boltzmann equation (\ref{ne10a}) by an arbitrary function $\Psi(\bx,\bv,t)$ and integrate the resulting equation by taking into account the invariant element of integration (\ref{ne10d}). The post-Newtonian version of the Maxwell-Enskog transfer equation, reads
\ben\no
\frac{\partial}{\partial t}\int\Psi\bigg[1+\frac1{c^2}\left(\frac{5v^2}2+7U\right)\bigg]fd^3v
+\frac{\partial}{\partial x^i}\int\Psi v_i\bigg[1+\frac1{c^2}\bigg(\frac{5v^2}2+7U\bigg)\bigg]fd^3v
\\\no
+\frac2{c^2}\int\Psi\bigg[\frac{\partial U}{\partial t}+\frac{\partial U}{\partial x^i}v_i\bigg] fd^3v
-\int\bigg[\frac{\partial\Psi}{\partial t}+\frac{\partial\Psi}{\partial x^i}v_i\bigg]\bigg[1+\frac1{c^2}\left(\frac{5v^2}2+7U\right)\bigg]fd^3v
\\\no
-\frac{\partial U}{\partial x^i}\int\frac{\partial\Psi}{\partial v^i}\left[1+\frac1{c^2}\left(\frac72v^2+3U\right)\right]fd^3v+\frac1{c^2}\int\frac{\partial\Psi}{\partial v^i}
\bigg[4v_iv_j\frac{\partial U}{\partial x^j}+3v_i\frac{\partial U}{\partial t}
\\\lb{ne14a}
-v_j\bigg(\frac{\partial \Pi_i}{\partial x^j}-\frac{\partial \Pi_j}{\partial x^i}\bigg)-2\frac{\partial \Phi}{\partial x^i}
-\frac{\partial \Pi_i}{\partial t}\bigg]fd^3v
=\int\Psi\left\{1+\frac1{c^2}\left[2v^2+6U\right]\right\}\mathcal{Q}(f,f)d^3v.
\een

Now the hydrodynamic equations can be obtained from the transfer equation (\ref{ne14a}) by choosing values of the arbitrary function  $\Psi(\bx,\bv,t)$ and integration of the resulting equations.

We begin by determining the mass density balance equation and for that end we choose $\Psi=m^4$ in (\ref{ne14a}), take into account  Grad's distribution function (\ref{ne13a}) and perform the integrations. Hence it follows
\ben\no
\frac{\partial}{\partial t}\bigg\{\rho\bigg[1+\frac1{c^2}\bigg(\frac{V^2}2+U\bigg)\bigg]\bigg\}+\frac{\partial}{\partial x^i}\bigg\{\rho V_i\bigg[1+\frac1{c^2}\bigg(\frac{V^2}2+U\bigg)\bigg]\bigg\}=-2\frac\rho{c^2}\bigg(\frac{\partial U }{\partial t}+V_i\frac{\partial U}{\partial x^i}\bigg)
\\\lb{ne14b}
=-\frac{2}{c^2}\bigg(\frac{\partial \rho U}{\partial t}+\frac{\partial \rho U V_i}{\partial x^i}\bigg)
+\frac{2U}{c^2}\underline{\bigg(\frac{\partial\rho}{\partial t}+
\frac{\partial \rho V_i}{\partial x^i}\bigg)},
\een
by rearranging the last term. For the underlined term we can use the Newtonian approximation of the continuity equation
\ben\lb{ne14c}
\frac{\partial\rho}{\partial t}+
\frac{\partial \rho V_i}{\partial x^i}=0.
\een
If we introduce in (\ref{ne14b})  the post-Newtonian  mass density \cite{Ch1,Fock}
\ben\lb{ne14d}
\rho^*=\rho\left[1+\frac1{c^2}\left(\frac{V^2}2+3U\right)\right],
\een
we get the continuity equation for the mass density $\rho^*$ in the post-Newtonian approximation
\ben\lb{ne14e}
\frac{\partial\rho^*}{\partial t}+\frac{\partial \rho^*V_i}{\partial x^i}=0.
\een
As it should be, the  above equation corresponds to the post-Newtonian  balance equation of the mass density for a perfect fluid \cite{Ch1,Fock}.

The mass-energy density hydrodynamic equation is obtained by choosing $\Psi=m^4u^0$ in the transfer equation (\ref{ne14a}) by considering  Grad's distribution function (\ref{ne13a}) and by integrating the resulting equation, yielding
\ben\no
\frac{\partial}{\partial t}\bigg\{\rho\bigg[1+\frac1{c^2}\left(V^2+2U+\frac{3kT}{2m}\right)\bigg]\bigg\}+\frac{\partial }{\partial x^i}\bigg\{\rho V_i\bigg[1+\frac1{c^2}\bigg(V^2+2U+\frac{5kT}{2m}\bigg)\bigg]\bigg\}
\\\lb{ne15a}
+\frac\rho{c^2}\frac{\partial U}{\partial t}+\frac1{c^2}\left(\frac{\partial \mathfrak{p}_{ij}V_j}{\partial x^i}+\frac{\partial \mathfrak{q}_i}{\partial x^i}\right)=0.
\een
Here we follow Chandrasekhar \cite{Ch1} and introduce the expression for the post-Newtonian mass-energy density
\ben
\sigma =\rho\left[1+\frac1{c^2}\left(V^2+2U+\frac{5kT}{2m}\right)\right]=\rho\left[1+\frac1{c^2}\left(V^2+2U+\varepsilon+\frac{p}\rho\right)\right],
\een
so that (\ref{ne15a}) can be rewritten as
\ben\lb{ne15b}
\frac{\partial\sigma}{\partial t}+\frac{\partial \sigma V_i }{\partial x^i}+\frac1{c^2}\left(\rho\frac{\partial U}{\partial t}-\frac{\partial p}{\partial t}\right)+\frac1{c^2}\left(\frac{\partial \mathfrak{p}_{ij}V_j}{\partial x^i}+\frac{\partial \mathfrak{q}_i}{\partial x^i}\right)=0.
\een
Without the new contributions of the pressure deviator $\mathfrak{p}_{ij}$ and heat flux $\mathfrak{q}_i$  (\ref{ne15b}) corresponds to the balance equation for the mass-energy density of a perfect fluid in the post-Newtonian approximation \cite{Ch1}.

For the hydrodynamic equation of the momentum density we choose $\Psi=m^4u^i$ in the transfer equation (\ref{ne14a}) use Grad's distribution function (\ref{ne13a}), integrate the resulting equation and get
\ben\no
\frac{\partial\sigma V_i}{\partial t}+\frac{\partial\sigma V_i V_j}{\partial x^j}+\frac{\partial}{\partial x^i}\left[p\left(1-\frac{2U}{c^2}\right)\right]+\frac{\partial\mathfrak{p}_{ij}}{\partial x^j}
-\rho\frac{\partial U}{\partial x^i}\bigg[1+\frac1{c^2}\left(2V^2+\varepsilon-2U-\frac{p}\rho\right)\bigg]
\\\no
+4\frac{\mathfrak{p}_{ij}}{c^2}\frac{\partial U}{\partial x^j}-\frac\rho{c^2}V_j\left(\frac{\partial \Pi_i}{\partial x^j}-\frac{\partial \Pi_j}{\partial x^i}\right)+4\frac\rho{c^2}V_i\left(\frac{\partial U}{\partial t}+V_j\frac{\partial U}{\partial x^j}\right)
-\frac\rho{c^2}\left(2\frac{\partial \Phi}{\partial x^i}+\frac{\partial \Pi_i}{\partial t}\right)
\\\lb{ne16a}
-\frac1{c^2}\frac{\partial\left(\mathfrak{p}_{ij}V_j+\mathfrak{q}_i\right)}{\partial t}\frac{\partial U}{\partial x^i}+\frac1{c^2}\frac{\partial}{\partial x^j}\bigg[\mathfrak{q}_iV_j+\mathfrak{q}_jV_i+\frac{\left(\mathfrak{p}_{ik}V_j+\mathfrak{p}_{jk}V_i\right)V_k}{2}\bigg]=0.
\een
Without the dissipative terms $\mathfrak{p}_{ik}$ and $\mathfrak{q}_i$ the above equation corresponds to eq. \emph{(68)} of \cite{Ch1}.

The  momentum density hydrodynamic equation (\ref{ne16a})  can be rewritten by taking into account the mass-energy hydrodynamic equation (\ref{ne15b}) as
\ben\no
\rho\frac{dV_i}{dt}+\frac{\partial\left(\mathfrak{p}_{ij}+p\delta_{ij}\right)}{\partial x^j}\left[1-
\frac1{c^2}\left(V^2+4U+\varepsilon+\frac{p}\rho\right)\right]-\rho\frac{\partial U}{\partial x^i}\left[1+\frac1{c^2}\left(V^2-4U\right)\right]
\\\no
+\frac\rho{c^2}\left[V_i\left(\frac1\rho\frac{\partial p}{\partial t}-\frac{\partial U}{\partial t}+4\frac{dU}{dt}\right)-2\frac{\partial \Phi}{\partial x^i}-\frac{d\Pi_i}{dt}+V_j\frac{\partial \Pi_j}{\partial x^i}
\right]+\frac1{c^2}\frac{\partial\left(\mathfrak{p}_{ij}V_j+\mathfrak{q}_i\right)}{\partial t}-2\frac{U}{c^2}\frac{\partial\mathfrak{p}_{ij}}{\partial x^j}
\\\lb{ne16b}
+\frac1{c^2}\frac{\partial}{\partial x^j}\left[\mathfrak{q}_iV_j+\mathfrak{q}_jV_i+\frac12\left(\mathfrak{p}_{ik}V_j+\mathfrak{p}_{jk}V_i\right)V_k+4U\mathfrak{p}_{ij}\right]
-\frac{V_i}{c^2}\left(\frac{\partial\mathfrak{p}_{jk}V_k}{\partial x^j}+\frac{\partial\mathfrak{q}_j}{\partial x^j}\right)=0.
\een
Here we have introduced the material time derivative $d/dt=\partial/\partial t+V_i\partial/\partial x^i$.

 By neglecting all terms of $\mathcal{O}\left(c^{-2}\right)$ order it follows the Newtonian momentum density  hydrodynamic equation
\ben\lb{ne16c}
\rho\frac{d V_i}{dt}+\frac{\partial \left(\mathfrak{p}_{ij}+p\delta_{ij}\right)}{\partial x^j}-\rho\frac{\partial U}{\partial x^i}=0.
\een

From the subtraction of the  mass density hydrodynamic equation  (\ref{ne14e}) from the mass-energy hydrodynamic equation (\ref{ne15b}) one can obtain the total energy density hydrodynamic equation, which is a sum of the internal $\rho\varepsilon$ and kinetic $\rho V^2/2$ energy densities, namely
\ben\no
\frac1{c^2}\bigg\{\frac{\partial}{\partial t}\left[\rho\left(\frac{V^2}2+\varepsilon\right)\right]+\frac{\partial}{\partial x^i}\left[\rho\left(\frac{V^2}2+\varepsilon\right)V_i\right]-\rho V_i\frac{\partial U}{\partial x^i}
+\frac{\partial\left[\mathfrak{p}_{ij}V_j+\mathfrak{q}_i+pV_i\right]}{\partial x^i}
\\\lb{ne17a}
+U\underline{\left(\frac{\partial\rho}{\partial t}+\frac{\partial\rho V_i}{\partial x^i}\right)}\bigg\}=0.
\een
Here we note that this equation is of order $\mathcal{O}(c^{-2})$ and we may use  the Newtonian continuity equation (\ref{ne14c}) for the underlined term so that (\ref{ne17a})  reduces to the Newtonian total energy density hydrodynamic equation for a viscous and heat conducting fluid
\ben\lb{ne17b}
\frac{\partial}{\partial t}\left[\rho\left(\frac{V^2}2+\varepsilon\right)\right]+\frac{\partial}{\partial x^i}\left[\rho\left(\frac{V^2}2+\varepsilon\right)V_i\right]-\rho V_i\frac{\partial U}{\partial x^i}
+\frac{\partial\left[\mathfrak{p}_{ij}V_j+\mathfrak{q}_i+pV_i\right]}{\partial x^i}
=0.
\een
We call attention to the fact that the post-Newtonian contributions to this equation do not show up. As was pointed out by Chandrasekhar \cite{Ch1,Ch2} the first post-Newtonian contributions to the total energy density  are obtained from  the knowledge of the second post-Newtonian contributions to the mass-energy hydrodynamic equation. In order to obtain these contributions here we have to determine Grad's distribution function in the second post-Newtonian approximation which is a heavy task and will be subject of a future work. The second post-Newtonian approximation to the Boltzmann equation and for the Maxwell-J\"uttner distribution function were determined in \cite{PGMK,GGKK}.

The internal energy density hydrodynamic equation follows from the elimination of the time derivative of the hydrodynamic velocity from (\ref{ne17b})  by using the Newtonian momentum density hydrodynamic equation (\ref{ne16c}), yielding
\ben\lb{ne17c}
\rho\frac{d\varepsilon}{dt}+\frac{\partial \mathfrak{q}_i}{\partial x^i}+\left[\mathfrak{p}_{ij}+p\delta_{ij}\right]\frac{\partial V_i}{\partial x^j}=0.
\een
This equation refers to the well-known  Newtonian internal energy density hydrodynamic equation for a viscous and heat-conducting
fluid.

The hydrodynamic equations derived from the transfer equation (\ref{ne14a}) for the mass, mass-energy and momentum densities are the same as those which follows  from the conservation laws for the particle four-flow (\ref{ne1})$_1$ and energy-momentum tensor (\ref{ne1})$_2$ (see \cite{GGKK}). 

Although the dynamic pressure appears in Grad's distribution function (\ref{ne13a}) it does not participate  in the Newtonian and first post-Newtonian  hydrodynamic equations. Furthermore, the collision term in the transfer equation vanishes, since mass, momentum and energy of a particle are conservative quantities at collision.

\section{Conclusions}\lb{sec5}

In this work the post-Newtonian energy-momentum tensor for a non-perfect gas described by the Eckart decomposition was determined. From the post-Newtonian Boltzmann equation a transfer equation was derived as well as the post-Newtonian expression for the relativistic Grad's distribution function. From the knowledge of the post-Newtonian transfer equation and Grad's distribution function the hydrodynamic equations for the mass density, mass-energy density and momentum density were determined which show the contributions of the viscous stress and heat conduction. The non-relativistic limiting case of these equations lead to the well-known Newtonian hydrodynamic equations for the mass, momentum and energy densities.

\section*{Appendix}
For the integration of the equations in the previous sections we have used the following well-known integrals from the kinetic theory of gases (see e.g. \cite{GK})
\ben\no
I_n=\int \mathcal{V}^ne^{- \frac{m\mathcal{V}^2}{kT}}d\mathcal{V}=\frac12\Gamma\left(\frac{n+1}2\right)\left(\frac{kT}m\right)^\frac{n+1}2
\qquad\Gamma(n+1)=n\Gamma(n),\quad \Gamma(1)=1, \quad \Gamma\left(\frac12\right)=\sqrt\pi,
\\\no
\int e^{- \frac{m\mathcal{V}^2}{kT_0}}\mathcal{V}_i\mathcal{V}_jd^3\mathcal{V}=\frac{I_{2}}{3} \delta_{ij},
\qquad\int \!e^{- \frac{m\mathcal{V}^2}{kT_0}}\mathcal{V}_i\mathcal{V}_j\mathcal{V}_k\mathcal{V}_ld^3\mathcal{V}=\!\frac{I_{4}\big[\delta_{ij}\delta_{kl}+ \delta_{ik}\delta_{jl}+\delta_{il}\delta_{jk}\big]}{15},
\\\no
\int e^{- \frac{m\mathcal{V}^2}{kT_0}}\mathcal{V}_i\mathcal{V}_j\mathcal{V}_k\mathcal{V}_l\mathcal{V}_n\mathcal{V}_nd^3\mathcal{V}
=\frac{I_{6}}{105}\big[\delta_{ij}\big(\delta_{kl}\delta_{mn} + \delta_{km}\delta_{ln}
+\delta_{kn}\delta_{lm}\big)+ \delta_{ik}\big(\delta_{jl}\delta_{mn}
+\delta_{jm}\delta_{ln}+ \delta_{jn}\delta_{lm}\big)
\\\no
+\delta_{il}\big(\delta_{jk}\delta_{mn}+ \delta_{jm}\delta_{kn}+\delta_{jn}\delta_{km}\big)+ \delta_{im}\big(\delta_{jk}\delta_{ln}
 +\delta_{jl}\delta_{kn}
 + \delta_{jn}\delta_{kl}\big)+
\delta_{in}\big(\delta_{jk}\delta_{lm}+\delta_{jl}\delta_{km}+\delta_{jm}\delta_{kl}\big)\big].
\een

%For a full discussion of the  role of the pressure deviator in the special relativistic hydrodynamic equations one is referred to the paper \cite{JH}.

\acknowledgments{ This work was supported by Conselho Nacional de Desenvolvimento Cient\'{i}fico e Tecnol\'{o}gico (CNPq), grant No.  304054/2019-4.}


\begin{thebibliography}{99}
\bibitem{Eins} A. Einstein, L. Infeld and B. Hoffmann,
The gravitational equations and the problem of motion, \emph{Ann. of Math.} {\bf39}, 65 (1938).
\bibitem{Ch1} S. Chandrasekhar, The post-Newtonian equations of hydrodynamics in general relativity, \emph{Ap. J.} {\bf142}, 1488 (1965).
\bibitem{Wein}  S. Weinberg, {\it Gravitation and cosmology. Principles and
applications of the theory of relativity} (Wiley, New York, 1972).
\bibitem{ChNu} S. Chandrasekhar and Y. Nutku, The second post-Newtonian equations of hydrodynamics in general relativity, \emph{Ap. J.} {\bf158}, 55 (1969).
\bibitem{PGMK} G.M. Kremer, Post-Newtonian  kinetic theory, {\it
Annals of Physics} {\bf 426}, 168400 (2021).
\bibitem{GGKK} G. M. Kremer, \emph{Post-Newtonian hydrodynamics: theory and applications}, to be published by Cambridge Scholars Publishing.
\bibitem{JH} J.-C. Hwang and H. Noh, Special relativistic hydrodynamics with gravitation, \emph{Ap. J.} {\bf833}, 180 (2016).
\bibitem{Eck} C. Eckart, The thermodynamics of irreversible processes, III. Relativistic theory of a simple fluid, \emph{Phys. Rev.} {\bf58}, 919 (1940).
\bibitem{CK} C. Cercignani and G. M. Kremer, \emph{The relativistic Boltzmann equation: theory and applications} (Birkh\"auser, Basel, 2002).
\bibitem{KRW} G. M. Kremer, M. G. Richarte and K. Weber, Self-gravitating systems of ideal gases in the 1PN approximation, Phys. Rev. D {\bf93}, 064073 (2016).
\bibitem{Grad} H. Grad, On the kinetic theory of rarefied gases, {\it Commun. Pure Appl. Math.} {\bf 2}, 331-407 (1949).
\bibitem{Ped} C. A. Ag\'on, J. F. Pedraza and J. Ramos-Caro, Kinetic theory of collisionless self-gravitating gases: Post-Newtonian polytropes, \emph{Phys. Rev. D} {\bf83}, 123007 (2011).
\bibitem{Rez} V. Rezania and Y. Sobouti, Liouville's equation in post Newtonian approximation I. Static solutions, \emph{Astron. Astrophys.} {\bf354}, 1110 (2000).


\bibitem{Max} J. C. Maxwell, On the dynamical theory of gases, \emph{Phil. Trans. R. Soc. London} {\bf 157}, 49 (1867).
\bibitem{Ens} D. Enskog, Bermerkungen zu einer Fundamentalgleichung in der kinetischen Gastheorie, \emph{Phys. Z.} {\bf12}, 534 (1911).
\bibitem{LL1}  L. D. Landau and E. M. Lifshitz, {\it The classical
theory of fields}, 4th ed.  (Pergamon Press, Oxford, 1980).
\bibitem{AbSt} M. Abramowitz and I. A. Stegun, {\it Handbook of mathematical
functions} (Dover, New York , 1968).
\bibitem{Fock} V. Fock,  {\it The theory of space time and gravitation}
(Pergamon Press, London, 1959).
\bibitem{Ch2} S. Chandrasekhar, Conservation laws in general relativity and in the post-Newtonian approximations, \emph{Ap. J.} {\bf158}, 45 (1969).
\bibitem{GK} G. M. Kremer, \emph{An introduction to the Boltzmann equation and transport processes in gases} (Springer, Berlin, 2010).




\end{thebibliography}
\end{document}